\documentclass[12pt,a4paper]{article}
\usepackage{mycommand}
\usepackage{graphicx}
\begin{document}
\textwidth=135mm
 \textheight=200mm
\begin{center}
{\bfseries Twist-3 quark-gluon correlation contribution to single-transverse spin
 asymmetry for direct-photon and single-jet productions in $pp$ collision
}
\vskip 5mm
K. Kanazawa$^{\dag}$ and Y. Koike$^\ddag$
\vskip 5mm
{\small {\it $^\dag$ Graduate School of Science and Technology, Niigata
 University, Ikarashi, Niigata 950-2111, Japan}} \\
{\small {\it $^\ddag$ Department of Physics, Niigata
 University, Ikarashi, Niigata 950-2111, Japan}}
\\
\end{center}
\vskip 5mm
\centerline{\bf Abstract}
We study the contribution of the collinear twist-3 quark-gluon
correlation to the 
single transverse-spin asymmetry in 
direct-photon and single-jet productions in $pp$ collision.
At typical RHIC kinematics 
we find sizable asymmetries in the forward region of the polarized nucleon while they are almost zero in
the backward region. We also find the soft-fermion
pole contribution
is vanishingly small,
suggesting the
measurement of these asymmetries provides us with an unique opportunity to determine the net
soft-gluon pole function.
\vskip 10mm
The study of large single transverse-spin asymmetry (SSA) in
inclusive reactions has
provided us with a range of new insights into the partonic structure of
hadrons.
%
When the transverse momentum of the final-state particle $P_T$ is large, the SSA can
be described as a twist-3
observable
in the framework of the collinear factorization~\cite{QiuSterman1992,EguchiKoikeTanaka2007,BeppuKoikeTanakaYoshida2010}.
In principle, the relevant twist-3 multiparton correlation effects 
exist 
both in the initial-state and the final-state hadrons.
In the inclusive hadron production
in $pp$ collision and semi-inclusive DIS, this coexistence makes 
it difficult to identify the origin of SSA uniquely.
The SSA for direct-photon and single-jet productions
play a crucial
role since the responsible effect exists only in
the initial-state hadrons.

In this report, as a reference for future RHIC experiment, we present
 our estimate of 
 the SSA in these processes, $A_N^{\gamma, \jet} = \Delta\sigma^{\gamma,
 \jet}
 /\sigma^{\gamma,
\jet}$, by taking into account the whole 
twist-3 quark-gluon correlation contribution which are composed of the
 soft-gluon pole (SGP) and soft-fermion pole (SFP) components.
A comparison with forthcoming measurements would shed new
 light on the net 
initial-state quark-gluon correlation inside the transversely polarized
 nucleon.

The relevant single-spin-dependent cross sections are schematically
represented as 
\cite{KouvarisQiuVogelsangYuan2006,KoikeTomita2009,KanazawaKoike2011DY}
\begin{eqnarray}
\Delta\sigma^{\gamma, \jet}
&\sim& \sum_{a,b} \left( G^a_{F}(x,x)-x\frac{dG^a_{F}(x,x)}{dx} \right) \otimes f^b(x')
\otimes \hat{\sigma}^{\rm SGP}_{ab\to \gamma, {\rm jet}} \nn\\
&+& \sum_{a,b} \left( G^a_{F}(0,x)+\widetilde{G}^a_{F}(0,x) \right)
\otimes f^b(x') \otimes
\hat{\sigma}^{\rm SFP}_{ab\to {\gamma, {\rm jet}}} , \label{qgc}
\end{eqnarray}
where the symbol $\otimes$ denotes the
convolution with respect to the partonic momentum fractions.
$\{ G_F^a,\GFt^a \}$ is a complete set of the quark-gluon correlation
function for flavor $a$\footnote{For the definition and property of
these functions, see
\cite{EguchiKoikeTanaka2006}.}, $f^b(x')$ is the unpolarized parton
distribution for flavor $b$, 
and $\hat{\sigma}_{ab\to\gamma,\jet}^{\rm SGP, SFP}$ are the corresponding partonic
hard cross sections.
The SGP and SFP functions, $G_F^a(x,x)$ and $G_F^a (0,x) + \GFt^a (0,x)$, for light-quark flavors
($a=u,d,s,\bar{u},\bar{d},\bar{s}$) have been determined by an analysis
of RHIC $A_N$ data for the inclusive pion
and kaon productions~\cite{KanazawaKoike2010}.
For the unpolarized parton distribution, we
use the GRV98LO parton distribution~\cite{GlueckReyaVogt1998} to keep
the consistency with our previous works.
The scales in the parton distribution and
fragmentation functions are taken as $\mu=P_T$.

\begin{figure}
 \begin{center}
  \fig{0.8}{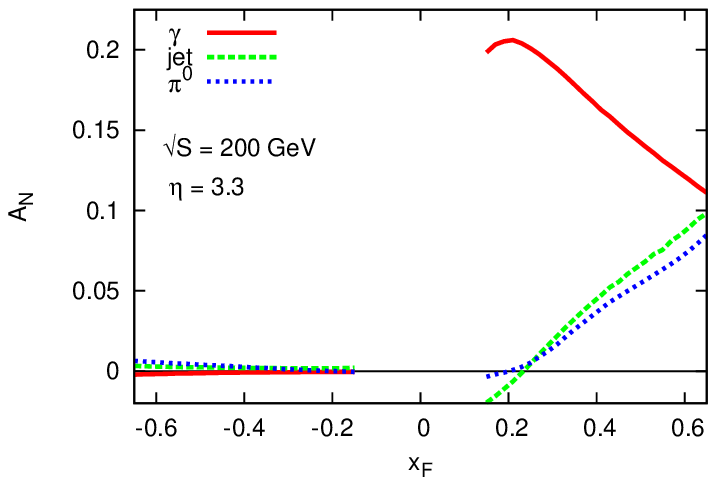}
  \fig{0.8}{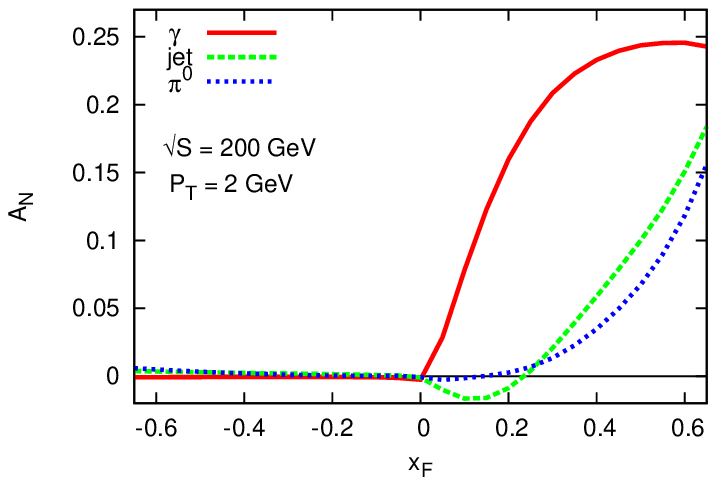}
 \end{center}
 \caption{Comparison of the $x_F$-dependence of $A_N$ for direct-photon,
 jet  and $\pi^0$ at fixed $\eta=3.3$ (left) and
 $P_T=2\GeV$ (right), respectively. The
 center-of-mass
 energy $\sqrt{S}$ is taken at 200 GeV. The plots are restricted in the
 region $P_T \ge 1 \GeV$.  \label{200}}
\end{figure}

\begin{figure}[t]
 \begin{center}
  \fig{0.8}{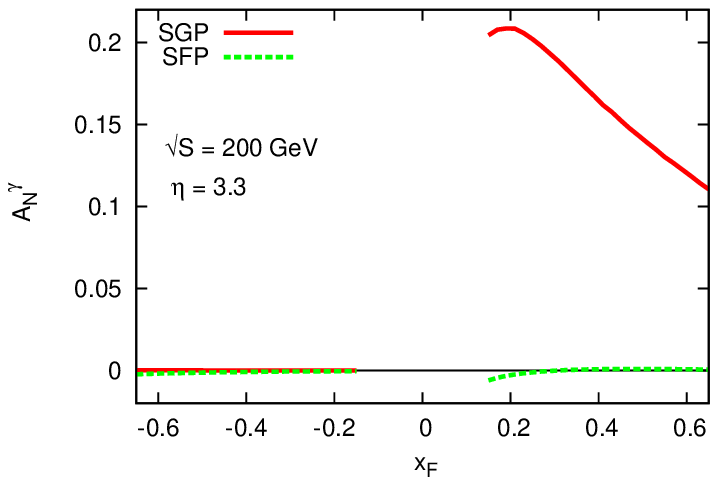}
  \fig{0.8}{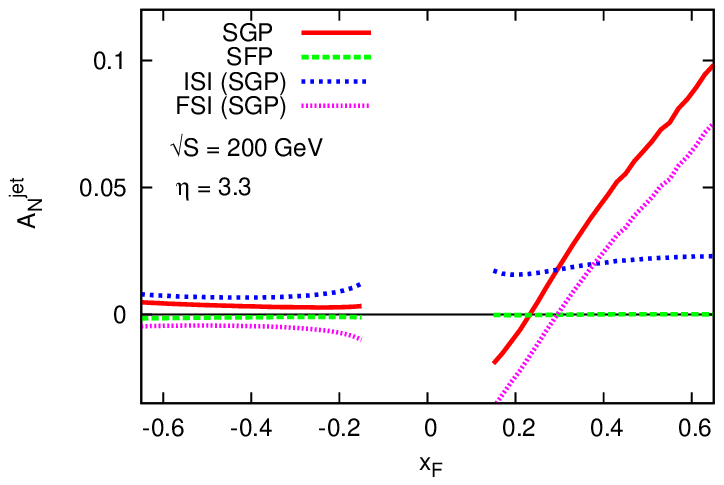}
 \end{center}
 \caption{Separation of $A_N^\gamma$ and $A_N^\jet$ in the left panel of
 Fig.~\ref{200} into the SGP and SFP
 components. Also
 shown in the right panel is the further separation of the SGP component into
 the initial-state and final-state interaction contributions.\label{bunri}}
\end{figure}

Figure \ref{200} shows the calculated $A_N^\gamma$ and $A_N^{\rm jet}$ 
as a function of Feynman-$x$ ($x_F$) at $\sqrt{S}=200\GeV$ for pseudorapidity $\eta$ and $P_T$, respectively, in comparison with
$A_N^{\pi^0}$.
From this figure one can see $A_N^{\gamma}$ is particularly large
and drops as a function of
$x_F$ in the forward region while $A_N^{\rm jet}$ behaves similarly to
$A_N^{\pi^0}$. 
These behaviors are qualitatively consistent with the previous
calculation\,\cite{KangQiuVogelsangYuan2011}.
One of the important features is that the large asymmetries arise only in
the forward region while they are almost zero in the backward region.
This observation is
interesting especially for the direct-photon process, because 
the 3-gluon correlation effect, which is another
source of SSA, could appear only in the
backward region~\cite{KoikeYoshida2012}. 
Namely, the role of the quark-gluon and 3-gluon
correlations for $A_N^\gamma$ is completely separated into the forward and backward
regions, respectively, so that
the measurement of $A_N^{\gamma}$ in whole
$x_F$-region could give a clear
information for both twist-3 multiparton correlations inside the
transversely polarized nucleon.

To see the relative importance of each pole contribution, we
show the separation of $A_N$ into the SGP and SFP
components in Fig.\,\ref{bunri}.
From this decomposition, one finds the large forward $A_N$ is mostly from the
SGP component while the SFP one is vanishingly small in whole $x_F$
region.
As shown in~\cite{KanazawaKoike2011}, in the case of
light-hadron production, a sizable SFP
contribution to $A_N$
manifests itself at moderate $x_F$ due to
the large gluon-to-pion and gluon-to-kaon fragmentation functions of the DSS
parametrization~\cite{FlorianSassotStratmann2007PIK}. 
For direct-photon and jet processes, however, due to the absence of the
large fragmentation function, 
the SFP contribution is sufficiently suppressed.
This suggests the SSA in these two processes would be golden
channels to probe the net SGP component of the quark-gluon correlation
function.
%

%
%

In summary, we have estimated the contribution of the twist-3 quark-gluon
correlation in the polarized nucleon for the SSA in the direct-photon
and the single-jet
productions at the typical RHIC kinematics. 
We have found $A_N^{\gamma}$ is particularly large in the forward region of the polarized
nucleon while the behavior of $A_N^{\rm jet}$ is similar to that for $\pi^0$.
In both processes, the SGP component
gives rise to dominant contribution for the forward SSA while the SFP
one is vanishingly small in the 
whole $x_F$-region. 
In addition, for the 
direct-photon process, we have shown there is no competition between the quark-gluon and 
the 3-gluon correlation contributions.
Owing to these 
features, the measurement of the forward $A_N^\gamma$ and $A_N^{\rm jet}$ would
provide an
unique opportunity for determining the SGP function in global QCD
analysis of SSA.


This work is supported by the Grand-in-Aid for
Scientific Research (No. 24.6959 and No. 23540292) from the Japan Society
of Promotion of Science.


\end{document}